\documentclass[twocolumn,showpacs,preprintnumbers,amsmath,amssymb]{revtex4}
\usepackage{graphicx}
\begin{document}

\title{'Electron' and 'photon' emerging
from supersymmetric neutral particles: A possible realization in
ultracold Bose-Fermi atom mixture}

\author{Yue Yu$^{1,2}$ and S. T. Chui$^2$}
\affiliation{1. Institute of Theoretical Physics, Chinese Academy
of Sciences, P.O. Box 2735, Beijing 100080, China}
 \affiliation{2. Bartol
Research Institute, University of Delaware, Newark, DE 19716, USA}
\date{\today}
\begin{abstract}
We show that the 'electron' and 'photon' can emerge from a
supersymmetric Hubbard model which is a non-relativistic theory of
the neutral particles. The Higgs boson and 'photon' may not appear
in the same phase of the phase diagram. In a Mott insulator phase
of the boson, the 'electron' and 'photon' are stablized by an
induced Coulomb interaction between 'electrons'. This emergent
mechanism may be 'realized' in an ultracold Bose-Fermi atom
mixture except the long range Coulomb interaction is repalced by a
nearest neighbor one. We suggest to create 'external electric
field' so that the 'electron' excitation can be observed by
measuring the linear density-density response of the 'electron'
gas to the 'external field' in the time flying experiment  of the
mixture. The Fermi surface of the 'electron' gas may also be
expected to be observed in the time flying.
\end{abstract}


\maketitle

What roles did the mysterious Higgs boson play except providing a
mass to the intermediate bosons in the weak-electrodynamics? Where
did the electron and photon come from \cite{wen}? In this Letter,
we suggest that it was a neutral world in high temperature and the
electrons and photons  emerged when the temperature was lower than
certain critical temperature. The Higgs boson fluctuation plays an
important role in the appearance of the finite gauge coupling
constant. We show that this emergence can be 'realized' by the
elementary excitations in the ultracold Bose-Fermi atom mixture in
optical lattices except the Coulomb interaction between
'electrons' is replaced by a nearest neighbor one.

We start from a three-dimensional spinless supersymmetric(SUSY)
Hubbard model, which was called 'ultracold superstring' in
one-dimensional lattice \cite{stoof}. The microscopic description
of the model parameters for an ultracold atom mixture has been
proposed \cite{aie}. The constitution particles in this model are
neutral spinless bosons ($a_i$) and fermions ($f_i$) with $i$ the
lattice site index. In high temperatures, there are only
excitations of these constitution particles. We call this a
confinement phase. After the SUSY is broken, the system undergoes
a phase transition, in certain critical temperature, to  a uniform
mean field (UMF) state in which a finite coupling gauge field and
a 'charged' fermion emerge if the fermion occupation is nearly
half filling. We call these emergent objects  'photon' and
'electron', respectively. At the exact half filling, this UMF
state turns to a long range ordered state, the checkerboard
crystal \cite{comp}. However, this UMF state can only be stable if
the lattice filling of the constitution bosons is integer and has
a Mott insulator (MI) ground state. For a special space structure,
the 'electron' interaction is pure Coulomb.

Experimentally, the Bose-Fermi atom mixture in optical lattices
has been realized for $^{87}$Rb-$^{40}$K \cite{rbk}, and
$^{23}$Na-$^6$Li \cite{nali}. We suggest an experiment to create
an 'external field' by changing the depth of the fermion's optical
potential. The response function of the mixture to the external
field may be measured by the density distribution in a time flying
of the cold atom cloud. The behavior of the response functions may
be used to identify the fermionic elementary excitations in a set
of given parameters, either the constitution one or 'electronic'.
On the other hand, we expect the 'electron' Fermi surface can be
observed by the time flying, which has been used to observe the
Fermi surface of the pure cold Fermi atoms \cite{kms}.

The Bose-Fermi mixture Hubbard Hamiltonian we are interested in
reads
\begin{eqnarray}
H&=&-\sum_{i\ne j}(t_{B,ij}a^\dagger_ia_j+t_{F,ij}f^\dag_i
f_j)-\sum_i (\mu_Bn^a_i+\mu_Fn^f_i)\nonumber\\
&+&\frac{U_{BB}}2\sum_i n^a_i(n^a_i-1)+U_{BF}\sum_i n^a_in^f_i,
\label{sh}
\end{eqnarray}
where the lattice spacing is set to unit. $n^a_i=a^\dag_ia_i$ and
$n^f_i=f^\dag_if_i$. $t_{B,ij}$ and $t_{F,ij}$ are the hopping
amplitudes of the boson and fermion between a pair of  sites.
$\mu_B$ and $\mu_F$ are chemical potentials. And $U_{BB}$ and
$U_{BF}$ are the on-site interactions between bosons, and between
boson and fermion, both of which may be repulsive or attractive.
The microscopic calculations of these model parameters in terms of
the cold atom mixture  may be found, e.g, in Ref. \cite{aie}. If
$t_{B,ij}=t_{F,ij}=t_{ij}$, $\mu_B=\mu_F=\mu$ and
$U_{BB}=U_{BF}=U$, the model is SUSY under transformations $
\delta a_i=i\theta f_i,~~~\delta f_i=i\theta a_i$ where $\theta$
is a Grassmann number.

To deduce the low energy theory in strong couplings , we will use
the slave particle technique, which has been applied to the cold
boson system \cite{d,yu,lu}. In the slave particle language, the
Hamiltonian reads $H=H_2+H_4$ where $H_2$ and $H_4$ are the
two-operator and four-operator terms, respectively. Namely,
\begin{eqnarray}
&&H_2=-\sum_i\sum_\alpha[\mu_B\alpha( n^\alpha_{c,i}+
 n^\alpha_{h,i})+\mu_F n^\alpha_{c,i}]\\
&&+\frac{U_{BB}}{2}\sum_i\sum_\alpha\alpha(\alpha-1)
(n_{c,i}^\alpha+n_{h,i}^\alpha)+U_{BF}\sum_i\sum_\alpha \alpha
n_{c,i}^\alpha,\nonumber
\end{eqnarray}
and
\begin{eqnarray}
H_4&=&-\sum_{
ij}t_{F,ij}\sum_{\alpha,\beta}c^\dagger_{\alpha,i}h_{\alpha,i}
h^\dagger_{\beta,j}c_{\beta,j}\nonumber\\&-&\sum_{
ij}t_{B,ij}\sum_{\alpha,\beta}\sqrt{\alpha+1}\sqrt{\beta+1}\\
&(&h^\dagger_{\alpha+1,i}h_{\alpha,i}+c^\dagger_{\alpha+1,i}c_{\alpha,i})
(h^\dagger_{\beta,j} h_{\beta+1,j}+c^\dagger_{\beta,j}
c_{\beta+1,j}),\nonumber
\end{eqnarray}
where $n^\alpha_{c,i}=c^\dagger_{\alpha,i}c_{\alpha,i}$ and
$n^\alpha_{h,i}=h^\dagger_{\alpha,i}h_{\alpha,i}$. We explain
briefly the deduction of this slave particle Hamiltonian. The
state configurations at an arbitrary given site consists of $\{
|\alpha,s\rangle~|~\alpha=0,1,2,...~;~s=0,1 \}$ where $\alpha$ and
$s$ are the boson and fermion occupations, respectively. The Bose
and Fermi creation operators can be expressed as
$a^\dag=\sum_\alpha\sqrt{\alpha+1}[|\alpha+1,0\rangle\langle
\alpha,0|+|\alpha+1,1\rangle\langle \alpha,1|], f^\dag=\sum_\alpha
|\alpha,1\rangle\langle\alpha,0|$. The mapping to the slave
particle reads $|\alpha,0\rangle\to h^\dag_\alpha,
|\alpha,1\rangle\to c^\dag_\alpha$ . We call $c_\alpha$ the
composite fermion (CF) \cite{comp} and $h_\alpha$ the slave boson.
The normalized condition
$\sum_\alpha(|\alpha,0\rangle\langle\alpha,0|+|\alpha,1\rangle\langle\alpha,1|)=1$
implies a constraint $\sum_\alpha(n^\alpha_{c}+n^\alpha_{h})=1$ at
each site. The slave particles arise a $U(1)$ gauge symmetry $
 c_{\alpha,i}\to
e^{i\varphi_i}c_{\alpha,i},h_{\alpha,i}\to
e^{i\varphi_i}h_{\alpha,i}. $ The SUSY transformation is given by
$ \delta h_{\alpha,i}=i\theta
\sqrt{\alpha}c_{\alpha-1,i},~~~\delta
c_{\alpha,i}=i\theta\sqrt{\alpha+1}h_{\alpha+1,i}$. The global
$U(1)$ symmetry $ c_{\alpha,i}\to e^{i\alpha\varphi}c_{\alpha,i},
h_{\alpha,i}\to e^{i\alpha\varphi}h_{\alpha,i}$ reflects the
particle number conservation. Since the slave particle technique
essentially works in the strong coupling region, we focus on $\bar
U_{BB}=U_{BB}/|\varepsilon_0|\gg 1$ where $\varepsilon_k$ is the
dispersion of the constitution boson. (Hereafter, a quantity $\bar
A\equiv A/|\varepsilon_0|$.)

There are two types of four slave particle terms in $H_4$, the
$t_B$ terms and $t_F$ terms. We first neglect the $t_B$ terms in
the mean field level of the CF. To decouple the $t_F$ terms , we
introduce Hubbard-Stratonovich fields
$\hat\eta^{c,h}_{\alpha\beta,ij}$ and
$\hat\chi^{c,h}_{\alpha\beta,ij}$. The partition function is given
by $Z=\int D\hat\chi D\hat\eta D\bar c Dc D\bar h Dh D\lambda
~e^{-S_{eff}}$ where the effective action reads
\begin{eqnarray}
&&S_{eff}[\hat\chi,\hat\eta,c,h,\lambda]=\int_0^{1/T} d\tau
\biggl\{\sum_i\sum_{\alpha}\nonumber\\&&\biggl[\bar
c_{\alpha,i}(\partial_\tau
-(\mu_B\alpha+\mu_F)+\frac{U_{BB}}2\alpha(\alpha-1)\nonumber\\
&&+U_{BF}\alpha-i\lambda_i)c_{\alpha,i}+\bar
h_{\alpha,i}(\partial_\tau
-\alpha\mu_B\nonumber\\
&&+\frac{U_{BB}}2\alpha(\alpha-1)-i\lambda_i)h_{\alpha,i}\biggr]
+\sum_ii\lambda_i\nonumber\\
&&+\sum_{ ij;\alpha,\beta}t_{F,ij}[\hat\eta^h_{\beta\alpha,ji}
(\hat\chi^c_{\alpha\beta,ij}-\bar
c_{\alpha,i}c_{\beta,j})-\hat\chi^c_{\alpha\beta,ij}\hat\chi^h_{\beta\alpha,ji}\nonumber\\
&&+\hat\eta^c_{\alpha\beta,ij}(\hat\chi^h_{\beta\alpha,ji} -\bar
h_{\beta,j}h_{\alpha,i})]+t_B~{\rm terms}\biggr\},
\end{eqnarray}
 where $\lambda_i$ is the Lagrange multiplier for
the constraint $\sum_\alpha(n^\alpha_{c,i}+n^\alpha_{h,i})=1$.
Rewriting
$\hat\eta^{c,h}_{\alpha\beta,ij}=\eta^{c,h}_{\alpha\beta,ij}e^{i{\cal
A}_{ij}}$ and
$\hat\chi^{c,h}_{\alpha\beta,ij}=\chi^{c,h}_{\alpha\beta,ij}e^{i{\cal
A}_{ij}}$ and $\lambda_i=\lambda+{\cal A}_{0,i}$, ${\cal A}_{0i}$
and ${\cal A}_{ij}$ are $U(1)$ gauge field corresponding to the
gauge symmetry. Fixing a gauge, $\hat
\eta_{\alpha\beta,ij}^h\approx
\eta^h_{\alpha,ij}\delta_{\alpha\beta}$, $\hat
\chi_{\alpha\beta,ij}^h\approx
\chi^h_{\alpha,ij}\delta_{\alpha\beta}$ and
$\lambda_i\approx\lambda$ (which is a saddle point value of
$\lambda_i$). This is corresponding to a mean field approximation.
The effective mean field action is given by
\begin{eqnarray}
S_{MF}&=&iN_s\beta\lambda+\int d\tau\biggl\{\sum_{\langle
ij\rangle}(
t_{F,ij}\sum_\alpha\chi^c_{\alpha,ij}\chi^h_{\alpha,ji})\\
&+&\sum_{i\ne j}\sum_{\alpha}(\bar
c_{\alpha,i}(D^\alpha_c)^{-1}_{ij}c_{\alpha,j}+\bar
h_{\alpha,i}(D^\alpha_h)^{-1}_{ij}h_{\alpha,j})\biggr\}\nonumber
\end{eqnarray}
where $(D^\alpha_c)^{-1}_{ij}=(\partial_\tau
-\mu^\alpha_F)\delta_{ij}
-t_F\sum_{\vec\tau}\chi^h_{\alpha,ji}\delta_{j,i+\vec \tau}$,
$\mu^\alpha_F=\mu_B\alpha+\mu_F
+\frac{U_{BB}}2\alpha(\alpha-1)+U_{BF}\alpha-i\lambda$ and
$(D^\alpha_h)^{-1}_{ij}=(\partial_\tau -\alpha\mu_B
+\frac{U_{BB}}2\alpha(\alpha-1)-i\lambda))\delta_{ij}-t_F\sum_{\vec\tau}\chi^c_{\alpha,ji}\delta_{j,i+\vec
\tau}$. ($\vec \tau$ is the unit vector of the lattice.) To
simplify, we have assumed only the nearest neighbor hopping $t_F$
of the fermions is not zero.

The mean field equations are $ \chi^c_{\alpha,ij}=\langle
c^\dag_{\alpha,i} c_{\alpha,j}\rangle=T\sum_n
D^\alpha_{c,ij}(p_n), \chi^h_{\alpha,ij}=\langle h^\dag_{\alpha,i}
h_{\alpha,j}\rangle=T\sum_n D^\alpha_{h,ij}(\omega_n)$ where
$D^\alpha_{c,ij}(p_n)=\int d\tau
e^{ip_n\tau}D^\alpha_{c,ij}(\tau)$ and
$D^\alpha_{h,ij}(\omega_n)=\int d\tau
e^{i\omega_n\tau}D^\alpha_{h,ij}(\tau)$ with $p_n$ and $\omega_n$
the Fermi and Bose frequencies. Solving the mean field equations
near the critical point with $\chi^\alpha\approx 0$, one can
obtain the critical temperature $T_c^\alpha$
\begin{eqnarray}
T_c^{\alpha}=\sqrt{T^\alpha_FT^\alpha_B}
\end{eqnarray}
 with
$ T^\alpha_{F,B}=t_Fn^\alpha_{c,h}(1\mp n^\alpha_{c,h})$, $
T^\alpha_c\chi^c_{\alpha,ij}=-T^\alpha_F\chi^h_{\alpha,ji}$ and
$T^\alpha_c\chi^h_{\alpha,ij}=-T^\alpha_B\chi^c_{\alpha,ji}$ where
$n_h^\alpha=1/[e^{\beta[-i\lambda-\alpha\mu_B+\alpha(\alpha-1)U_{BB}/2]}-
1],n_c^\alpha=1/[e^{\beta[-i\lambda-(\alpha\mu_B+\mu_F)
+\alpha(\alpha-1)U_{BB}/2+\alpha U_{BF}]}+ 1]$. Below this
temperature, the minimized free energy including variables $\chi$
is $ F\propto -t_F\sum_\alpha(\tau^\alpha
)^2\frac{(S^\alpha_2)^2}{S^\alpha_4}$  where
$\tau^\alpha=\frac{T^\alpha_c-T}T$ and $S^\alpha_2$, $S^\alpha_4$
are the fractions of non-zero terms corresponding to the second
order and four order of $|\chi^\alpha|$ \cite{il}. The optimal
$|\chi^c_\alpha|^2\propto \tau^\alpha S^\alpha_2/S^\alpha_4$.

Before going to a concrete mean field state, we first require the
mixture is stable against the Bose-Fermi phase separation. For
example, it was known that the mixture is stable in the MI phase
if $4\pi t_F\sin(\pi n^f) U_{BB}>U_{BF}^2$ \cite{aie}. For the
parameters in the SUSY model, this inequality is not satisfied.
Thus, we assume the SUSY has been broken. In cold atom contents,
the SUSY is broken by hands through changing the lattice
potentials. If the real electron and photon are generated in the
way described in this paper, the SUSY breaking might be related to
the change of space-time structure in early universe, which has
been beyond the present model.

We consider an integer boson filling $n^a=$1 for $\bar U_{BB}\gg
1$. In this region, since $n^{\alpha\ne 1}_h$ and $n^{\alpha\ne
1}_c$ are very small, there are no solutions of the mean field
equations with $T^{\alpha\ne 1}_c\geq 0$ for the critical
temperature equations. Thus, all slave bosons and CFs with
$\alpha\ne 1$ are confined. Only $T^1_c>0$ can be found. Below
$T_c^1$, the CF $c_1$  and the slave boson $h_1$ are deconfined in
the mean field sense. We shall see the CF $c_1$ may be identified
as the 'electron' while $h_1$ as the Higgs boson. In the inset of
Fig. 1, we plot $T_c^1$ for a set of given parameters. The curve
$T_p(U_{BB})$ is corresponding to $\mu^1_F(T_p)=0$, i.e., the
effective chemical potential of $c_1$ vanishes. For the fermion
half filling, $\mu_F^1(T=0)\to 0$. This means that the Fermi
surface of CF $c_1$ shrinks and implies an insulating state, a
checkerboard crystal as we shall see soon. Meanwhile, we also find
$T_c^1=0$ for the half-filling of the fermions.

We now go to concrete mean field solutions and focus on the near
half filling of the fermions. There is no flux mean field state in
the present model because $\chi^{h,c}$ are real. Two possible
solutions are the dimer and uniform phase. The dimer phase may be
favorite in the fermion half-filling. However, sightly away from
the half filling, the most stable mean field state is the uniform
phase, as that for the $t$-$J$ model \cite{tj,il}, say, for
$\alpha=1$, each bond is endowed with the same real values of
$\chi^{c,h}_\alpha$ if a CF (i.e., $a_i^\dag
f_i^\dagger|0\rangle=c^\dag_{1i}|{\rm vac}\rangle$) is surrounded
by six bosons (i. e., $a_j^\dag|0\rangle=h^\dag_{1i}|{\rm
vac}\rangle$). In the fermion half filling, this is a checkerboard
crystal state \cite{comp}. Slightly away from the half filling,
this is a UMF state with $h_1$-$c_1$ bond, similar to that in the
fermion Hubbard model or $t$-$J$ model (see, e.g., \cite{wen}). In
the uniform phase, we choose $\chi^c_\alpha>0$. The dispersions of
the CFs and slave bosons are $ \xi_\alpha^{c,h}(k)=\mp
t_F|\chi_\alpha^{h,c}| \sum_i\cos k_i$. The CF ground state is
around ${\bf k}=0$ while the slave boson's is around ${\bf
k}={\vec \pi}$.

To see the stability of the UMF state, one has to add back the
gauge fluctuations on the mean field solution. The zeroth
component ${\cal A}_{i0}$ restores the exact constraint of one
slave particle per site. This fluctuation may unstabilize the mean
field state. Before integrating over $h$ or $c$, the system is in
the strong coupling limit because the gauge field has no dynamics,
i.e., the Lagrangian of the gauge field is $ L_{g}=-\lim_{g\to
\infty}\frac{1}{g}F^{\mu\nu}F_{\mu\nu}$ (in the continuous limit).
This means even the slave particles have a finite mass in
$T<T^\alpha_c$, the system is still in charge confinement phase.
Near $k=0$, one has to integrate away $h_1$. Then, the coupling
constant of the gauge field becomes finite. In the continuous
limit, the effective gauge action reads
\begin{eqnarray}
S[{\cal A}]=\sum_n\frac{T}2\int d^3k {\cal A}_\mu(k,\omega_n){\cal
A}_\nu(k,\omega_n)\Pi^B_{\mu\nu}(k,\omega_n),
\end{eqnarray}
where $\Pi^B_{\mu\nu}$ is the response function integrating over
$h_1$.

The long wave length behaviors of these response functions are
strongly dependent on the phase structure of the constitution
bosons. Introduce  a Hubbard-Stratonovich field
$\Phi_i=\sum_\alpha\sqrt{\alpha+1}[c^\dag_{\alpha+1,i}c_{\alpha,i}
+h^\dag_{\alpha+1,i}h_{\alpha,i}]$ to decouple $t_B$ terms
\cite{d,yu}. $\Phi_i$ may be thought as the order parameter field
of the Bose condensation. The phase boundary of the boson
superfluid (BSF)/normal liquid phases can be determined  by
$G^{-1}(0,0)=0$ \cite{d} where the $\Phi$-field propagator is
given by
\begin{eqnarray}
G^{-1}({\bf k},i\omega_n)&=&\varepsilon_{\bf k}-\varepsilon_{\bf
k}^2\sum_\alpha(\alpha+1)\biggl\{\frac{n_h^\alpha-n_h^{\alpha+1}}
{i\omega_n+\mu_B -\alpha U_{BB}} \nonumber\\
&+&\frac{n_c^\alpha-n_c^{\alpha+1}} {i\omega_n+\mu_B -\alpha
U_{BB}-U_{BF}}\biggr\}.
\end{eqnarray}
As expected, the phase diagram of the constitution boson consists
of the Bose superfluid (BSF), the normal liquid and the MI, in
which the MI phase only exists in the zero temperature and an
integer boson filling factor . In the inset of Fig. 1, we show an
example for the same parameters as the CF mean field phase
diagram.

We now go back to the response function $\Pi^B_{\mu\nu}$. The
current-current response function has the form
$\Pi^B_{ij}=(\delta_{ij}-k_ik_j/k^2)\Pi_L+(k_ik_j/k^2)\Pi_T$. In
the normal liquid and MI of the boson, the longitudinal part
$\Pi_L(0,0)=0$ and the gauge field is massless. If $h_1$
condenses, $\Pi_L(k,\omega) \to |\langle h^1\rangle|^2 \ne 0 $
when $k,\omega \to 0$. This gives the gauge field a mass. This is
well-known Higgs mechanism. In
 this sense, $h_1$ may be identified as the Higgs boson. This
 happens in the BSF.

\begin{figure}
\begin{center}
\includegraphics[width=7cm]{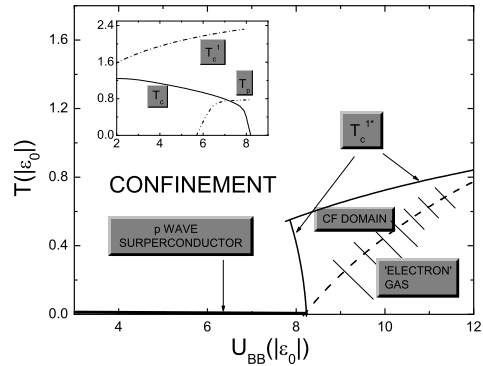}
\end{center}
 \caption{The phase diagram of the CFs in the $U_{BB}$-$T$ plane for
  $n^a=1$, $U_{BF}=U_{BB}/2$
 $\bar t_F=\sqrt{60}$, $n^f=0.55$. The solid curves
 are the critical temperature $T^{1*}_c$ after considering
 the gauge fluctuation and the shrink of the CF Fermi surface.
 The dash lines and the slashed area is the crossover
 from the 'electron' gas to the CF domain. The inset is the mean field
 phase diagram for both bosons and fermions.}
 \label{1}
\end{figure}

The density-density response function $\Pi^{B(0)}_{00}$ of free
slave boson yields an effective attraction between CFs in the long
wave length limit. This means that the UMF state is not stable if
$t_B$ term does not contribute to $\Pi^{B}_{00}$, which is indeed
the case in the BSF phase. In the MI, the $t_B$ term may be
thought as a perturbation if $U_{BB}$ and $|U_{BB}-U_{BF}|$ are
much larger than $t_B$. For the $n^a_i=1$ MI phase, the second
order perturbation and the MI character, $n^c_{1i}+n^h_{1i}=1$,
$t_B$ terms contribute an effective interaction between the CFs
 or slave bosons
\begin{eqnarray}
{\cal J}=\sum_{i\ne j}J_{ij}n^c_in^c_j+{\rm const}= \sum_{i\ne
j}J_{ij}n^h_in^h_j+{\rm const'},
\end{eqnarray}
with $J_{ij}=\frac{16 t_{B,ij}^2
U_{BF}^2}{U_{BB}(U_{BB}^2-U_{BF}^2)}$. We see that this
interaction is attractive if $U_{BF}>U_{BB}$ while it is repulsive
if $U_{BF}<U_{BB}$. For the repulsive one, if
$t_{B,ij}=t_B/\sqrt{|{\bf r}_i-{\bf r}_j|}$ for a special spatial
structure, the effective interaction between CF or the slave boson
in the continuous limit is a pure Coulomb one $ {\cal
J}=\sum_{k,k',q}V(q)c^\dagger_{{\bf k}+{\bf q}}c_{\bf k}
c^\dagger_{{\bf k'}-{\bf q}}c_{\bf
k'}=\sum_{k,k',q}V(q)h^\dagger_{{\bf k}+{\bf q}}h_{\bf k}
h^\dagger_{{\bf k'}-{\bf q}}h_{\bf k'}$ with
\begin{eqnarray}
V(q)=\frac{4\pi e^2}{q^2}, \label{coulomb}
\end{eqnarray}
and the 'charge' of the CF is defined by $e^2=\frac{16 t_{B}^2
U_{BF}^2}{U_{BB}(U_{BB}^2-U_{BF}^2)}$. This repulsive interaction
keeps the stability of the UMF state against the gauge
fluctuations. Eq. ({\ref{coulomb})  means that the CFs in the MI
phase behave like 'electrons'. 'Electrons' are a Coulomb gas.
Notice that the electron gas at the half filling ($n^f=1/2$) turns
to a checkerboard 'electron' lattice. In this way, we see that the
'electron' and 'photon' (the gauge field) emerge from a neutral
SUSY particle model.

In the ulrtacold Bose-Fermi mixture, instead of a pure Coulomb
interaction caused by a long range hopping $t_{B,ij}$, the
interaction between nearest neighbor sites dominates which is
caused by $t_{B,ij}=t_B\ne 0$ if only $(i,j)$ is a nearest
neighbor pair. The interaction between CFs at $T=0$ is given by
$V(q)=\frac{32\pi t_B^2U_{BF}^2}{U_{BB}(U_{BB}^2-U_{BF}^2)}$.
(Also see \cite{comp}.)

According to these discussions, we now may figure out the phase
diagram of the CF in Fig. 1. The mean field phase transition
temperature $T_c^1$ is suppressed greatly to $T_c^{1*}$ which is
determined by the $T_c$ and $T_p$. The dash line is the estimated
crossover line from the 'electron' gas (the MI of bosons) to  CF
domains which arises from the effective attraction between CFs. In
the BSF, the induced attraction between CFs,
$(\Pi^{B(0)}_{00}(q,0))^{-1}\sim-\frac{m_B q^2}{\rho_0}$, may lead
to a $p$ wave superconducting ground state. However, this might be
difficult to reach in the ultracold mixture \cite{you}.

 The experimental implications
of the UMF phase are discussed as follows. We consider the
'electron' response to an external 'electric' field, 'made' by a
change of the lattice potential of the fermions. This disturbs the
density of fermions  with $H'(t)=-\sum_i n^f({\bf
r}_i,t)\varphi({\bf r}_i,t)$. In a time flying experiment, the
difference between disturbed and undisturbed fermion densities by
external field is given by $n_f({\bf r})-n^0_f({\bf
r})=(\frac{m_f}t)|\tilde w_f({\bf k}=\frac{m_f{\bf r}}t)|^2\delta
n_f({\bf k}=\frac{m_f{\bf r}}t)$ where $t$ is the flying time,
$\tilde w_f$ is the Fourier component of the fermion Wannier
function and $\delta n_f({\bf k})\propto \Pi_{00}({\bf k},0)$ . If
$T>T_c^{1*}$, the density response of the system is simply given
by the free fermion one, $\Pi^{F(0)}_{00}$. For $T<T^{1*}_c$,
since $n^f\approx n_c^1$, the two CF response function is given by
$ \Pi_{00}^{-1}=(\Pi^{F(0)}_{00})^{-1}+(\Pi^{B(0)}_{00})^{-1}$.
Negative $\Pi^{B(0)}_{00}$ leads to the instability of the CF.
However, near the MI, the RPA gives $
\Pi_{00}^{-1}=(\Pi^{F(0)}_{00})^{-1}+(\Pi^{B(0)}_{00})^{-1}+V(q)$.
The repulsive between CFs stabilizes the CF against the gauge
fluctuation. A better experimentally measurable quantity is the
visibility ${\cal V}=\frac{n_f({\bf r}_{\rm max})-n_f({\bf r}_{\rm
min})}{n_f({\bf r}_{\rm max})+n_f({\bf r}_{\rm min})}$ where ${\bf
r}_{\rm max}$ and ${\bf r}_{\rm min}$ are chosen such that the
Wannier envelop is cancelled \cite{vi}. The difference between the
disturbed and undisturbed visibility may directly correspond to
the response function because $n_f({\bf r}_{\rm max})+n_f({\bf
r}_{\rm min})\approx n^0_f({\bf r}_{\rm max})+n^0_f({\bf r}_{\rm
min})$ in denominator.

The Fermi surface of pure cold fermions has been observed in a
recent experiment \cite{kms} by the time flying experiment. We
suggest to do the same observation to the mixture. It is expected
that instead of the constitution fermion's Fermi surface, one may
observe the 'electron' Fermi surface in the 'electron' gas.

In conclusions, we deduced the low energy physics of the SUSY
Hubbard model. The 'electron' and 'photon', as well as the Higgs
boson, were thought as emergent objects. The phase diagram was
depicted, which showed what happened as the temperature was from
high to low. Possible experiments to verify this theory were
suggested by means of the cold Bose-Fermi atom mixture.  We expect
to generalize our theory to include a non-abelian gauge field
\cite{wl} coupled to fermions with spin degrees of freedom.
However, it will not be a simple generalization of the $U(1)$
theory. Many problems, such as quark asymptotic free in an
ultraviolet limit and confinement at low energy, have to be
solved. The origin of the chiral fermions in $SU(2)\times U(1)$
weak-electric interaction is also non-trivial. We expect a
relativistic version of the present theory.

This work was supported in part by Chinese National Natural
Science Foundation and the NSF of USA.

\end{document}